# Structural transition and anisotropic properties of single crystalline SrFe$_2$As$_2$

J.-Q. Yan,[1] A. Kreyssig,[1,2] S. Nandi,[1,2] N. Ni,[1,2] S. L. Bud'ko,[1,2] A. Kracher,[1] R. J. McQueeney,[1,2] R. W. McCallum,[1,3] T. A. Lograsso,[1] A. I. Goldman,[1,2] and P. C. Canfield[1,2]

[1]Ames Laboratory, Ames, IA 50011

[2]Department of Physics and Astronomy, Iowa State University, Ames, IA 50011

[3]Materials Science and Engineering, Iowa State University, Ames, IA 50011

Plate-like single crystals of SrFe$_2$As$_2$ as large as 3×3×0.5 mm$^3$ have been grown out of Sn flux. The SrFe$_2$As$_2$ single crystals show a structural phase transition from a high temperature tetragonal phase to a low temperature orthorhombic phase at T$_o$ = 198 K, and do not show any sign of superconductivity down to 1.8 K. The structural transition is accompanied by an anomaly in the electrical resistivity, Hall resistivity, specific heat, and the anisotropic magnetic susceptibility. In an intermediate temperature range from 198 K to 160 K, single crystal X-ray diffraction suggests a coexistence of the high-temperature tetragonal and the low-temperature orthorhombic phases.

The recent discovery of high transition temperature ($T_c$) superconductivity in fluorine-doped RFeAsO (R = rare earth) and K-doped $AFe_2As_2$ (A = Sr and Ba) has attracted extensive attention in scientific community.[1, 2, 3] RFeAsO with the ZrCuSiAs-type structure and $AFe_2As_2$ with the $ThCr_2Si_2$-type structure share the same structural unit of ($Fe_2As_2$) layers. The ($Fe_2As_2$) layers are separated by ($R_2O_2$) layers in RFeAsO and by simple A-layers in $AFe_2As_2$. Partial substitution of $A^{2+}$ by $K^+$ or $Cs^+$ induces superconductivity with $T_c$ up to 38 K.[2, 3] Studies of polycrystalline samples have revealed phase transitions at $T_o$ = 140 K and 205 K for $BaFe_2As_2$ and $SrFe_2As_2$, respectively.[4,5] At $T_o$ = 140 K for $BaFe_2As_2$, the structural transition from a high temperature tetragonal phase to a low temperature orthorhombic phase is thought to be accompanied by a magnetic phase transition.[4] However, whether a similar structural transition takes place in $SrFe_2As_2$ at $T_o$ = 205 K concomitantly with the magnetic anomaly is still unknown. This question is of great importance due to the fact that a similar combination of a structural and magnetic phase transition was observed in RFeAsO.[6]

High quality, sizable single crystals are necessary in order to study the anisotropic properties of the compounds and details of the crystallographic structure. Ni et al. successfully grew $BaFe_2As_2$ single crystals out of Sn flux and studied the crystallographic, anisotropic magnetic and transport properties.[7] However, ~1% Sn was incorporated into the $BaFe_2As_2$ lattice and modifies the physical properties relative to those of polycrystalline samples. Whether $SrFe_2As_2$ lattice is susceptible to Sn incorporation remains to be clarified. Here, we report the growth of $SrFe_2As_2$ single crystals out of Sn flux and subsequent measurements of the anisotropic physical properties, including X-ray diffraction studies of the structural transition.

Single crystals of $SrFe_2As_2$ were grown out of Sn flux. Elemental Sr, Fe, and As were fired at 850°C for 12 hours and 900°C for 20 hours with intermediate grinding. The prefired pellet, which contains mainly $SrFe_2As_2$ with ~5% FeAs impurity, was broken into smaller pieces and added to Sn flux in the ratio of $[SrFe_2As_2]$:Sn = 1:48 and placed in a 2 ml crucible. A catch crucible containing quartz wool was mounted on top of growth crucible and both were sealed in a silica ampoule under approximately 1/3 atmosphere of argon gas. The packing and assembly of the growth ampoule was performed in air since the prefired pellet is stable in air. The sealed ampoule was heated to 1000°C and cooled over 36 hours to 500°C in a programmable furnace. At 500°C, the Sn was decanted from the $SrFe_2As_2$ crystals. We note that some pieces of the prefired pellets are still undissolved in the growth crucible. Raising the ratio of $[SrFe_2As_2]$:Sn leads to more residual undissolved $SrFe_2As_2$. Single crystals of $(Sr_{1-x}K_x)Fe_2As_2$ have also been grown using elemental starting components, as in [7], with similar results.

As-grown crystals are plate-like with a typical dimension of 1-2 mm × 1-2 mm × 0.1-0.5 mm. Some crystals manifest linear dimensions as large as 3-4 mm as shown in the inset of Figure 1. The crystallographic *c*-axis is perpendicular to the plane of the plate-like single crystals. Room temperature powder X-ray diffraction confirmed that the crystals are single phase with lattice parameters of a = 3.926(3) Å and c = 12.42 (1) Å, consistent with previous reports.[3,5] Elemental analysis of the crystals was performed using wavelength dispersive x-ray spectroscopy (WDS) in the electron probe microanalyzer (EPMA) of a JEOL JXA-8200 Superprobe. Elemental analysis confirmed the atomic ratio of 1:2:2 and observed about 0.3%(atomic) Sn in the bulk, a factor of three times less Sn than the $BaFe_2As_2$ crystals grown in a similar manner.[7]

Single crystal X-ray diffraction measurements were performed on a standard four-circle diffractometer using Cu $K_\alpha$ radiation from a rotating anode X-ray source, selected by a Ge(1 1 1) monochromator. For these measurements, a plate-like single crystal with dimensions of $3 \times 2 \times 0.5$ mm$^3$ was selected. The sample was mounted on a flat copper sample holder in a closed cycle displex cryogenic refrigerator with the (001)-(110) reciprocal lattice plane coincident with the scattering plane. The diffraction patterns were recorded for temperatures between 10 K and 300 K. The mosaicity of this crystal was 0.035° full width half maxima for the (0 0 10) reflection, indicating the excellent quality of the single crystal.

Magnetic properties were measured with a Quantum Design (QD) Magnetic Properties Measurement System (MPMS). The temperature dependent specific heat as well as magnetic field and temperature dependent electrical transport data were collected using a QD Physical Properties Measurement System (PPMS). Electrical contact was made to the samples using Epotek H20E silver epoxy to attach Pt wires in a 4-probe configuration. Hall measurements were performed in a QD PPMS instrument using the four probe ac (f = 16 Hz, I = 1 mA) technique with the current flowing in the ab plane approximately parallel to the a-axis and the field parallel to the c-axis. To eliminate the effect of a misalignment of the voltage contacts, the Hall measurements were taken for two opposite directions of the applied field, H and −H, and the odd component, $[\rho_H(H) - \rho_H(-H)]/2$, was taken as the Hall resistivity.

As shown in Figure 1, magnetic measurements clearly show the anisotropy in the magnetic susceptibility and a transition at $T_o = 198$ K. The transition temperature is slightly lower than that observed for polycrystalline samples.[5] Compared with the temperature dependence for polycrystalline samples, three prominent features are noteworthy: (1) Over the whole temperature range, there is a clear anisotropy with $\chi_{ab} > \chi_c$. (2) Below room temperature, both $\chi_{ab}$ and $\chi_c$ decrease linearly as the sample is cooled with a small slope until $T_o$. At $T_o$, they show a step-like change. Below $T_o$, both curves show a minimum in the temperature interval 100 K < T < 150 K. This temperature dependence is different from that of polycrystalline samples reported by Krellner et al.,[5] but agrees with that by Pfitsterer and Nagorsen.[8] (3) Below ~100 K the susceptibility becomes even more anisotropic, with a clear Curie-Weiss-like tail for H || ab. The effective moment associated with this tail is 0.35 $\mu_B$/f.u. The fact that this tail is so anisotropic argues that it is not associated with polycrystalline impurities within the residual Sn flux on the surface, but rather associated with the bulk, crystalline sample.

Figure 2(a) presents the temperature dependence of the in-plane electrical resistivity, $\rho(T)$. $\rho(T)$ decreases with cooling and shows a sudden drop at $T_o = 198$ K. The residual resistivity ratio (RRR) of $\rho_{300K}/\rho_{2K} \sim 6$ is smaller than 32 of polycrystalline samples[5] and larger than ~3 of similar single crystals grown out of Sn flux.[9] The application of a magnetic field of 140 kOe parallel to the c-axis has no effect on the phase transition at $T_o \sim 198$ K but slightly enhances the resistivity below ~150 K. On the other hand, the application of a 140 kOe field perpendicular to the c-axis shifts the transition upward by as much as ~10 K.

The inset of Figure 2 (a) shows the field dependence of the magnetoresistivity (MR) measured at 2 K. When the magnetic field is applied along the c-axis, the MR increases with field and reaches ~27% by 140 kOe. In contrast, the MR starts to increase only for H > 30 kOe and only reaches ~18% by 140 kOe when the magnetic field is applied in ab plane. The Hall resistivity (Figure 2b) is almost temperature independent at high temperatures, but has a sharp break just below 200 K, consistent with a structural phase transition. On further cooling, initially the Hall resistivity increases slightly, and then, below ~ 150 K monotonically decreases, reaching, at 2 K, a value close to that observed for $BaFe_2As_2$.[7]

The temperature dependent specific heat data are presented in Figure 3. A sharp feature at T = 198 K (lower inset) indicates the transition temperature which is consistent with that determined from magnetic and transport measurements. The upper inset to Figure 3 shows the low temperature $C_p/T$ data plotted as a function of $T^2$. Below $T^2 \sim 20$ $K^2$ the data deviate from linearity possibly associated with residual Sn flux. The fitting for $20 < T^2 < 45$ $K^2$ to the standard power law, $C_p = \gamma T + \beta T^3$ yields $\gamma = 33$ mJ/mol $K^2$ and $\beta = 0.64$ mJ/mol $K^4$ ($\Theta_D = 196$ K) similar to those of $BaFe_2As_2$.[7]

The transition temperature determined from the magnetization, electrical resistivity, Hall resistivity, and specific heat measurements agrees with each other. Comparable lattice parameters and similar transition temperatures for $SrFe_2As_2$ single and polycrystals are consistent with the WDS analysis that incorporated Sn is significantly less in $SrFe_2As_2$ single crystals than that in $BaFe_2As_2$ crystals. To determine whether there is a structural transition at $T_o$ as observed in the isostructural $BaFe_2As_2$,[4,7] we performed a single crystal x-ray diffraction study. As illustrated in Figure 4, below 160 K, the splitting of the (1 1 10) reflection was observed in ($\xi$ $\xi$ 0) scans, whereas the shape of the (0 0 10) reflection is unchanged in both ($\xi$ $\xi$ 0) and (0 0 $\zeta$) scans. This behavior is consistent with a tetragonal-to-orthorhombic phase transition with a distortion along the diagonal (1 1 0) direction and a transition from the space group *I4/mmm* to *Fmmm*, similar to that observed in the $BaFe_2As_2$ compound.[7] In an intermediate temperature range between 190 K and 160 K, reflections related to the orthorhombic phase appear to coexist with reflections related to the tetragonal phase as shown in the ($\xi$ $\xi$ 0) scans in Figure 4. However, no significant difference was observed between measurements with decreasing and increasing temperatures as might be expected from a hysteretic 1$^{st}$-order transition. We note that the split reflections are not equal in intensity or width at low temperatures, perhaps indicating an unequal 'twin domain' population. However, we also note that a crystallographic structure with symmetry lower than the orthorhombic can yield a similar diffraction pattern at low temperature and can not be excluded by the present study.

The lattice parameters have been determined by measurements of selected scans for the (0 0 10) and (1 1 10) reflections. By analyzing the position of the (0 0 10) reflection in longitudinal (0 0 $\zeta$) scans, the $c$-lattice parameter can be determined. The in-plane $a$ and $b$-lattice parameters have been calculated based on the distance between the reflections close to the tetragonal (1 1 10) position and the (0 0 10) reflection in transverse scans along the ($\xi$ $\xi$ 0) direction. The results are shown in Figure 5 together with an analysis of the intensity of the characteristic reflections. Between 10 K and 160 K, the difference in the orthorhombic $a$ and $b$-lattice parameters decreases monotonically with increasing temperature. Surprisingly, the difference between the in-plane lattice parameters stays nearly constant between 160 K and the transition to the tetragonal phase above 190 K. In the same intermediate temperature range, however, the sign of the thermal expansion coefficient for the orthorhombic **b** direction changes, coincident with the abrupt appearance of the central reflection related to the tetragonal phase at 160 K (see inset in Figure 5). These observations suggest a more complex nature for this phase transition even though the transition from the space group *I4/mmm* to *Fmmm* can be second order. We also note that while the structure in this intermediate temperature range can be described in terms of a coexistence between the high-temperature tetragonal phase and the low-temperature orthorhombic phase, a complex superstructure of the high-temperature phase with an additional order parameter could likewise yield the observed set of three reflections. It should be noted that measurements of M(T)/H on the crystal used for Figs. 4 and 5 showed only one transition at $T_o$ = 198 K.

In conclusion, we have successfully grown sizable single crystals of $SrFe_2As_2$ out of Sn flux and studied anisotropic thermodynamic and transport properties as well as the structural transition. Electrical resistivity, Hall resistivity, specific heat, and the anisotropic magnetic susceptibility demonstrate a transition at $T_o = 198$ K. The magnetic susceptibility is anisotropic, with $\chi\|ab > \chi\|c$ for 2 K < T < 300 K and, while the transition appears to be insensitive to $H\|c = 140$ kOe, it does shift upward by ~10K for $H\|ab = 140$ kOe. As evidenced by the single crystal x-ray diffraction study, a similar structural phase transition as in $BaFe_2As_2$ from a high temperature tetragonal phase to a low temperature orthorhombic phase takes place at $T_o = 198$ K. The observation of similar crystallographic phase transitions in $AFe_2As_2$ (A=Sr and Ba) and RFeAsO (R = rare earth) compounds suggests that the structure of the ($Fe_2As_2$) layers, present in all of these compounds, may play an important role in the superconductivity of the F- and K-doped analogues.

**Acknowledgment**

Work at the Ames Laboratory was supported by the US Department of Energy - Basic Energy Sciences under Contract No. DE-AC02-07CH11358. The authors would like to acknowledge useful discussions with E. D. Mun, M. Tillman, M. G. Kim, G. E. Rustan.

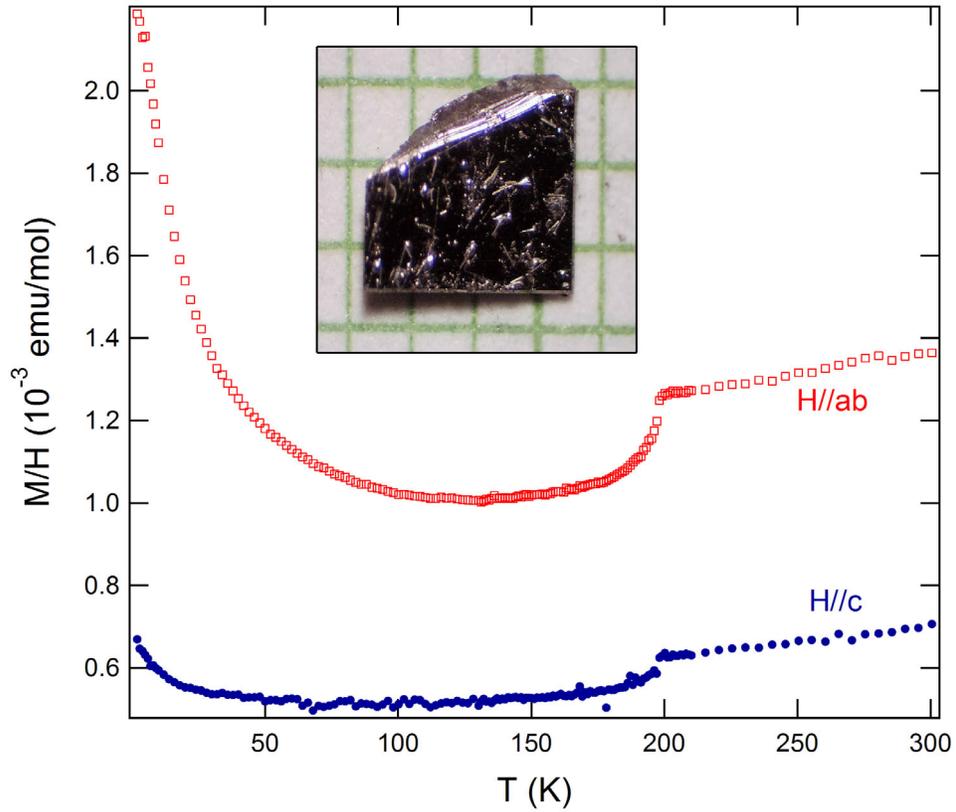

Fig. 1: Temperature dependent, anisotropic magnetization measured at a magnetic field of 50 kOe for $SrFe_2As_2$. Inset shows the photograph of a single crystal on a mm grid. The crystallographic c-axis is perpendicular to the plane of the plate. A few droplets of Sn flux can be seen on the surface.

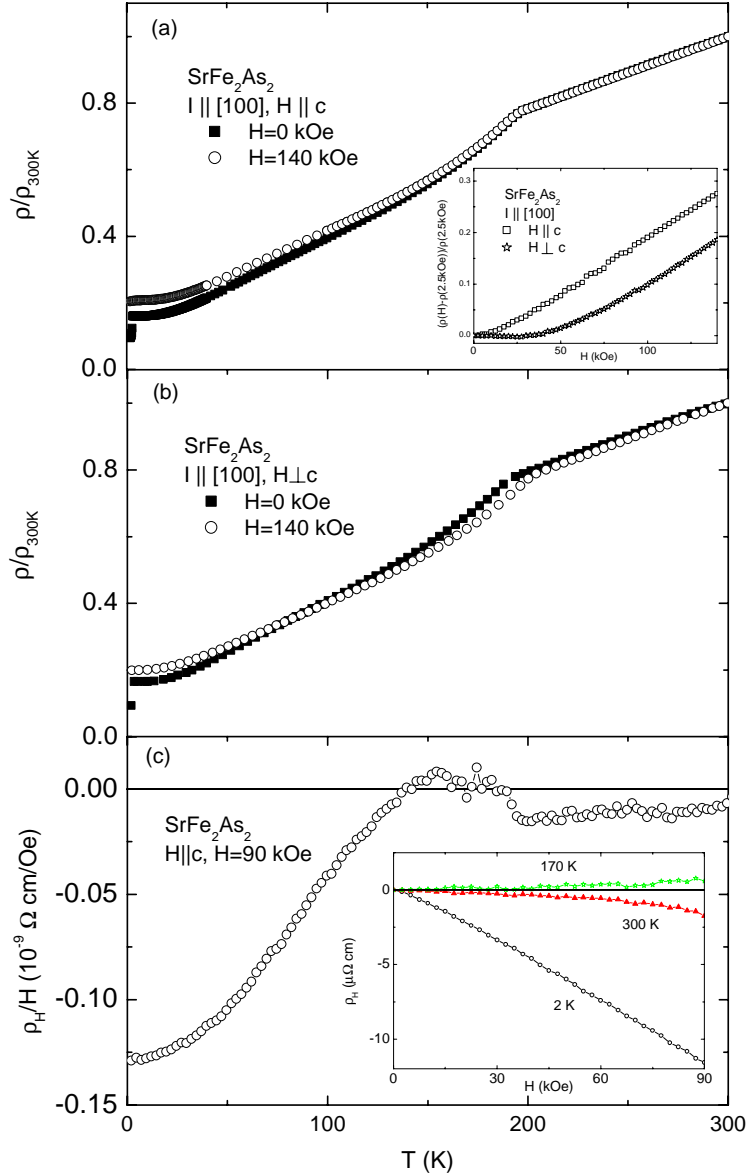

Fig 2: (a, b) Temperature dependence of normalized in-plane electrical resistivity in zero field and in an applied magnetic field of 140 kOe. Inset shows the field dependence of the magnetoresistivity $\Delta\rho/\rho_0$ measured at 2 K with magnetic field applied parallel and perpendicular to the c-axis. Note small features in the H=0 data at low temperatures are due to small amounts of Sn (flux) on the sample. (c) Temperature dependence of the Hall

resistivity divided by applied magnetic field. The inset shows a close to linear magnetic field dependence of the Hall resistivity at three salient temperatures: 300K, 170 K and 2 K.

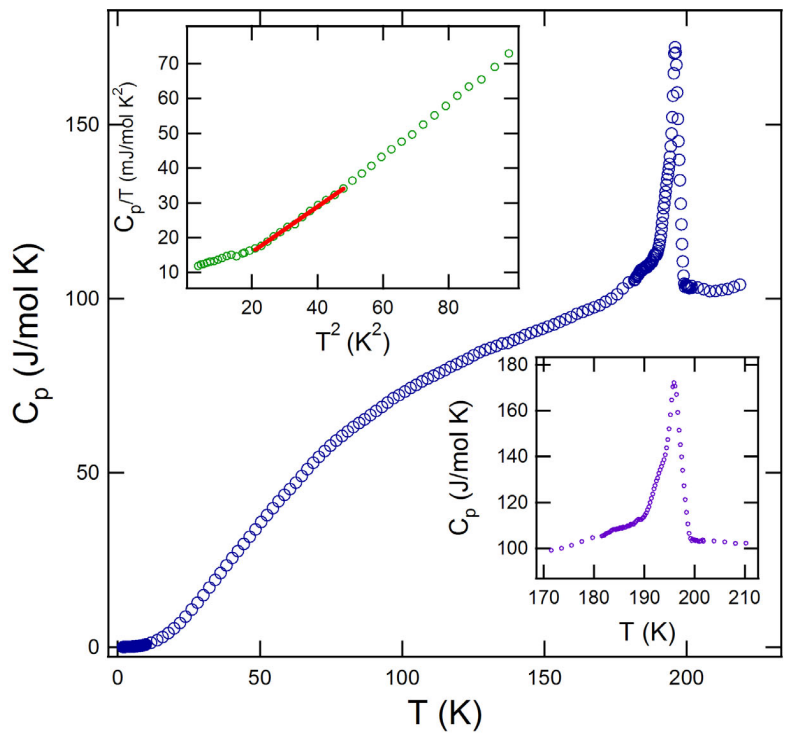

Fig 3 Temperature dependent specific heat of $SrFe_2As_2$. Upper inset: Cp/T as a function of $T^2$ for low temperature data. Lower inset: expanded view near the $T_o = 198$ K phase transition.

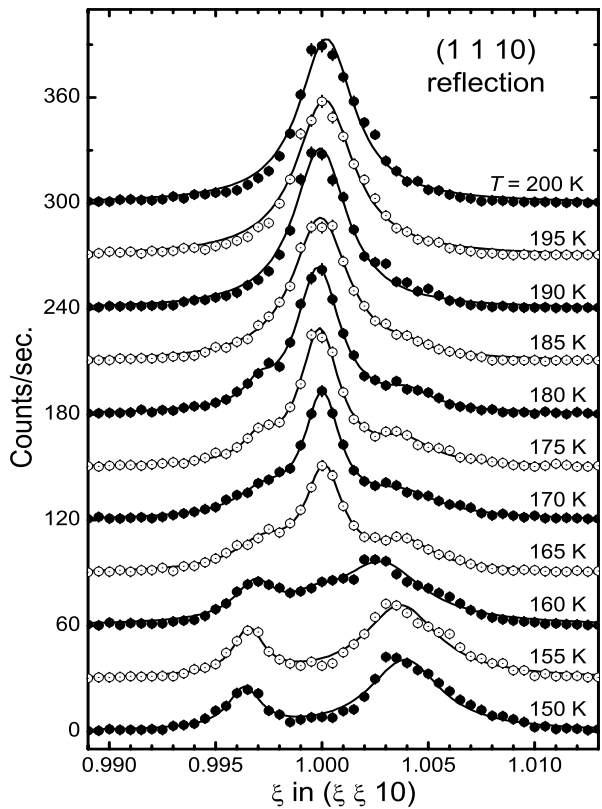

Fig 4: X-ray diffraction scans along the (1 1 0) direction through the position of the tetragonal (1 1 10) reflection for selected temperatures. The lines represent the fitted curves to obtain the reflection positions for the data shown in Figure 5. The offset between every data set is 30 Counts/sec.

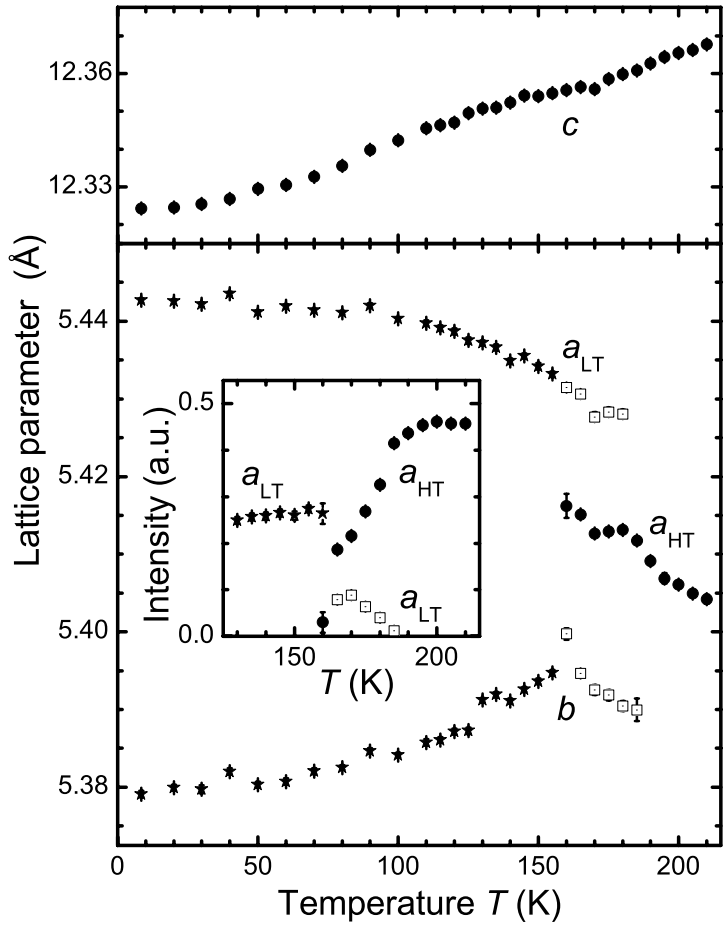

Fig 5: Lattice parameters for the tetragonal and orthorhombic phases as extracted from the data shown in Figure 4 for the (1 1 10) reflection and the (0 0 10) reflection (not shown). To allow direct comparison of both phases the lattice parameter $a$ of the tetragonal phase is given as $a_{HT}=\sqrt{2}a$. The inset shows the intensity of the reflections related to the lattice parameter $a$ in both phases. The error bars represent the relative precision, the absolute error is method-related and significantly larger.